%% file: ijcai26_main.tex
\documentclass{article}
\usepackage{ijcai26}

\usepackage{times}
\usepackage{soul}
\usepackage{url}
\usepackage[hidelinks]{hyperref}
\usepackage[utf8]{inputenc}
\usepackage[small]{caption}
\usepackage{amsmath}
\usepackage{amssymb}
\usepackage{amsthm}
\usepackage{bbm}
\usepackage{booktabs}
\usepackage{algorithm}
\usepackage{algorithmic}
\usepackage[switch]{lineno}
\usepackage{multirow}
\usepackage{array}
\usepackage{bbm}
\usepackage[table]{xcolor}
\usepackage{subcaption}
\usepackage{graphicx}
\usepackage[inline]{enumitem}
\usepackage{cleveref}

\colorlet{mycolor}{gray!10}

\title{Modeling Behavioral Intensity and Transitions for Generative Recommendation}

\author{
Wenxuan Yang$^{1,2}$\thanks{This work was conducted during the internship at Ant Group.}\and
Xiaoyang Xu$^1$\and\\
Hanyu Zhang$^2$\and
Zhexuan Xu$^1$\and
Wanqiang Xiong$^1$\and 
Zhaoqun Chen $^1$ \\
\affiliations
$^{1}$Ant Group\and
$^{2}$Fudan University\\
\emails
wenxuanyang2469@gmail.com,\\
\{xuxiaoyang.xxy, xuzhexuan.xzx, wanqiang.xwq, zhaoqun.czq\}@antgroup.com
hanyuzhang24@m.fudan.edu.cn
}

\begin{document}

\maketitle

\begin{abstract}
Multi-behavior recommendation aims to predict user conversions by modeling various interaction types that carry distinct intent signals. Recently, generative sequence modeling methods have emerged as an important paradigm for multi-behavior recommendation by achieving flexible sequence generation. However, existing generative methods typically treat behaviors as auxiliary token features and feed them into unified attention mechanisms. These models implicitly assume uniform activation of dependencies among historical behaviors, thereby failing to discern differences in intensity or capture transition patterns. To address these limitations, we propose BITRec, a novel generative multi-behavior recommendation framework that introduces structured behavioral modeling through selective dependency activation. BITRec incorporates (i) Hierarchical Behavior Aggregation (HBA), which explicitly models behavioral intensity differences through separated exploration and commitment pathways, and (ii) Transition Relation Encoding (TRE), which encodes transition structures through explicit learnable relation matrices. Experiments on four large-scale datasets (RetailRocket, Taobao, Tmall, Insurance Dataset) with millions of interactions achieve consistent improvements of 15-23\% across multiple metrics, with peak gains of 22.79\% MRR on Tmall and 17.83\% HR@10, 17.55\% NDCG@10 on Taobao.
\end{abstract}

\section{Introduction}

Recommendation systems capture users' decision-making processes through behavioral funnels: from exposure and clicks to add-to-cart and purchases. Different behaviors reflect qualitatively distinct user states—shallow behaviors indicate exploration, while deep behaviors signal explicit purchase intent. Accurately modeling this behavioral heterogeneity and structured progressions within funnels is crucial for predicting conversions, yet remains a core challenge.

Recently, multi-behavior generative recommendation models~\cite{MBgen,EAGER} have emerged as a promising end-to-end paradigm by leveraging autoregressive sequence modeling. Existing methods~\cite{hstu,hllm,pinrec} typically adopt one of three modeling sequence strategies: 
1) homogeneous behavior sequences, 2) interleaved item-behavior sequences, or 3) fused item-behavior representation sequences.
These methods then apply standard attention mechanisms~\cite{SASRec,hllm,onerec,MBgen}, to aggregate historical interactions, where dependencies are governed by content similarity (query-key matching) and temporal proximity (positional encoding). However, this uniform dependency aggregation overlooks two fundamental properties of multi-behavior sequences.

\begin{figure*}[t]
  \centering
\includegraphics[width=0.9\textwidth]{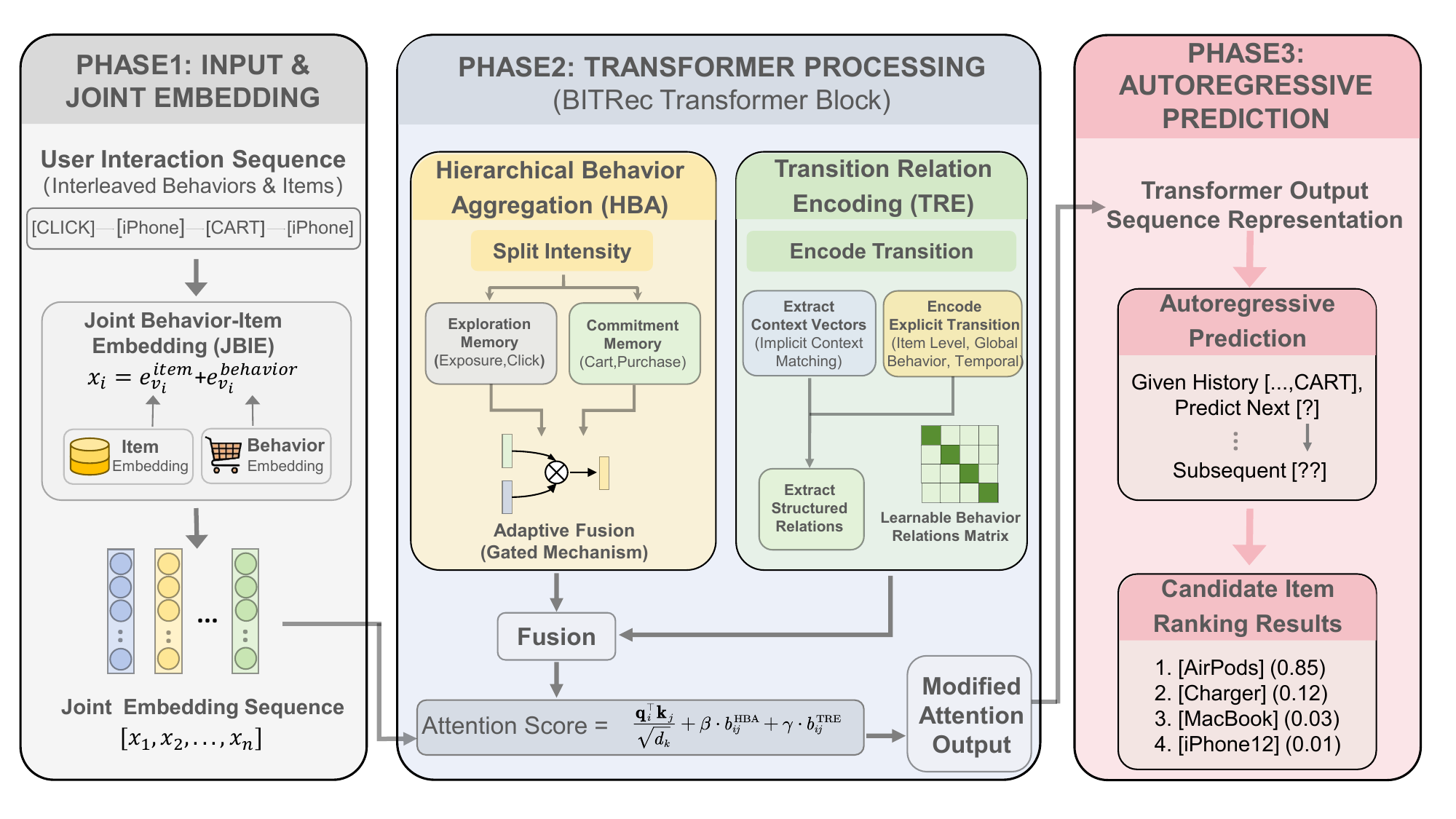}
\vspace{-15pt}
\caption{Overview of BITRec architecture. \textbf{Phase 1:} Joint Behavior-Item Embedding fuses behaviors and items into composite tokens. \textbf{Phase 2:} BITRec Transformer processes sequences via Hierarchical Behavior Aggregation (separating exploration/commitment representations) and Transition Relation Encoding. \textbf{Phase 3:} Autoregressive prediction generates recommendations.}
  \label{fig:framework}
  \vspace{-10pt}
\end{figure*}

First, behavioral intensity heterogeneity: different behaviors exhibit varying intensity levels. Shallow behaviors (such as clicks) reflect exploratory engagement, whereas deep behaviors (e.g. add-to-cart, favoriting) indicate stronger purchase intent. In e-commerce scenarios, users might casually click dozens of items yet ultimately add only one to cart; in high-commitment domains (e.g., insurance), this tendency is even more pronounced: users generate numerous page views without ever converting. Therefore, when predicting purchases, 20 casual clicks should not outweigh 1 deliberate add-to-cart. However, standard attention mechanisms \cite{MBgen,DMBIN,Tiger} aggregate these behaviors uniformly via softmax weights based on content similarity ignoring behavioral intensity altogether. Since intensity (e.g., clicks vs. add-to-cart) reflects user intent independently of content relevance, treating superficial signals on par with strong intent indicators overweights the exploration signals and introduces substantial noise, ultimately undermining predictive accuracy.

Second, directional transition structure: user trajectories exhibit pronounced funnel structures where purchases build upon exploratory behaviors (browsing, clicks, consultations). Traditional methods \cite{NMTR,mbht,PBAT} explicitly model this hierarchical dependency through carefully crafted fusion mechanisms, while existing generative methods compress this structure by injecting behavior-augmented tokens into standard attention. For instance, an add-to-cart from five minutes ago serves as a direct funnel predecessor, indicating strong purchase intent. Meanwhile, an item click occurred just one minute ago, despite temporally closer, usually represents an earlier exploration behavior. Standard attention relies on recency and content similarity, neglecting the directional progression and stage-dependent dynamics inherent in behavioral funnels, consequently limiting downstream conversion prediction.

The root cause lies in the uniform dependency assumption: existing methods assume dependencies are universally present and activated based solely on content similarity and temporal distance. In reality, dependencies should be selectively triggered based on what the user does (behavioral semantics) and how behaviors relate (transition structure).

To address these limitations, we propose BITRec (Behavioral Intensity and Transitions Generative Recommendation model) that introduces selective dependency activation into generative recommendation, where dependencies are conditionally triggered rather than uniformly aggregated. BITRec achieves this through two core mechanisms. Hierarchical Behavior Aggregation (HBA) decomposes sequences into exploration (exposure, clicks) and commitment (add-to-cart, purchase) representations, adaptively emphasizing commitment signals when predicting conversions. Transition Relation Encoding (TRE) learns behavior-specific relation matrices and structured features (item affinity, temporal proximity) to capture funnel-like dependencies. To enable these mechanisms, we adopt Joint Behavior-Item Embedding that fuses behaviors and items at the representation level.

Our main contributions include:
\begin{itemize}
\item We propose BITRec (Behavioral Intensity and Transitions for Generative Recommendation) that introduces the selective dependency activation framework, where dependencies are conditionally triggered based on behavioral semantics and transition structure.
\item We design Hierarchical Behavior Aggregation (HBA) to enable selective activation based on behavioral intensity, and Transition Relation Encoding (TRE) to enable selective activation based on transition patterns.
\item We validate BITRec on four diverse datasets, achieving state-of-the-art performance across all benchmarks and yielding substantial 18--23\% improvements on large-scale industrial scenarios (Taobao/Tmall).
\end{itemize}

\section{Related Work}

\subsection{Sequential Recommendation }
Sequential recommendation methods predict the next item by capturing dependencies within a user's historical interaction sequences. Among them, Transformer-based models stand out due to their powerful attention mechanism for sequence modeling, leading to the best performance in recommendation tasks \cite{DBLP:conf/sigir/HouHZZ22,zhang2019}. For example, SASRec \cite{SASRec} leverages unidirectional self-attention to model long-range dependencies inherent in user history sequences. BERT4Rec \cite{BERT4Rec} employs bidirectional self-attention to model user behavior sequences, allowing each item in the history to incorporate information from both its left and right contexts for recommendation. Despite their success, these models typically treat all interactions as uniform, failing to account for specific behavior types that are essential for understanding user intent.

\nocite{TDM}


\subsection{Multi-Behavior Recommendation }
In real-world scenarios, users interact with items through multiple behaviors. Multi-behavior recommendation aims to model these heterogeneous behaviors to mine fine-grained user intents \cite{abdel2013survey,zeng2019user,MB‑GMN,NMTR,DMBIN}.

\vspace{1pt}
\noindent
\textbf{Graph-based Multi-Behavior Modeling.}
Early multi-behavior approaches focused on static correlations between behaviors. MBGCN \cite{MBGCN} constructs a unified graph to represent multi-behavior data, learning behavior intensities through a user-item propagation layer and capturing behavior semantics through an item-item propagation layer. COPF \cite{copf} models multi-behavior fusion as a combinatorial optimization problem, applying stage-specific constraints to restrict the solution space and improve efficiency. These methods are effective in capturing associations, but typically rely on static graph structures and fail to fully capture the dynamic evolution of behaviors over time.



\vspace{1pt}
\noindent
\textbf{Sequence-based Multi-Behavior Modeling.}
Multi-behavior sequential models aim to model the heterogeneity and temporal dependencies inherent in user behavior. MBHT \cite{mbht} utilizes multi-scale temporal dynamics to capture behavior-aware item transitions, while employing a dynamic multi-order relational learning paradigm to model multi-behavioral dependencies. PBAT \cite{PBAT} models personalized patterns and multifaceted sequential collaborations to boost recommendation performance. 
Although effective, these discriminative models often have difficulty in modeling the joint probability of complex behavioral sequences.



\nocite{gamer}
\nocite{li2025blade}

\subsection{Generative Recommendation}
Recent works formulate recommendation as sequence generation. TIGER \cite{Tiger} pioneered generative retrieval via semantic IDs, while HSTU \cite{hstu} and HLLM \cite{hllm} verified LLM scaling laws and hierarchical architectures in this domain. However, these methods primarily focus on single interaction types, neglecting diverse behavioral patterns. To address this, recent studies incorporate multi-behavior data: MBGen \cite{MBgen} interleaves behavior and item tokens; GAMER \cite{gamer} models hierarchical dependencies via cross-level interactions; and BLADE \cite{li2025blade} employs dual item-behavior fusion. Despite these advances, current methods largely treat behavior tokens as auxiliary features within unified attention mechanisms, failing to explicitly discern behavior intensity or capture inherent transition structures.

\begin{table}[!t]
    \centering
    \caption{Statistics of different datasets.}
    \vspace{-8pt}
    \label{tab:dataset_statistics_transposed}
    
    \footnotesize
    
    \setlength{\tabcolsep}{1pt}
    
    \begin{tabular}{l cccc}
        \toprule
        \textbf{Metric} 
        & \textbf{RetailRocket}
        & \textbf{Taobao} 
        & \textbf{Tmall(10\%)} 
        & \textbf{Insurance} \\
        \midrule
        \# Users 
        & 1.4M & 988K & 977K & 1.3M \\
        
        \# Items 
        & 235K & 4.2M & 6.2M & 2.3K \\
        
        \# Interactions 
        & 2.8M & 100M & 133M & 14.2M \\
        
        Avg. Seq. Len. 
        & 1.96 & 101.4 & 136.7 & 10.6 \\
   
        Max Seq. Len.
        & 7.8K & 848 & 108K & 13.8K \\
        
        Behavior Types 
        & 3 & 4 & 4 & 6 \\
        \bottomrule
        \vspace{-15pt}
    \end{tabular}
\end{table}

\section{Methodology}
In this section, we present BITRec (Behavioral Intensity and Transitions for Generative Recommendation), a generative framework for multi-behavior sequential recommendation. The overall architecture is illustrated in Figure ~\ref{fig:framework}.

\subsection{Problem Formulation}
We formalize the multi-behavior sequential recommendation task as a generative problem. Let $\mathcal{U}$ and $\mathcal{V}$ denote the user set and item set, respectively, where each item $v \in \mathcal{V}$ is associated with a category $c \in \mathcal{C}$ (e.g., electronics, apparel, books). 

\vspace{2pt}
\noindent\textbf{Hierarchical Behavioral Structure.} User behaviors form a hierarchical structure $\mathcal{B} = \{b_0, b_1, \ldots, b_k\}$ where $b_i < b_j$ indicates lower engagement. This reflects the decision funnel: page-view ($b_0$), click ($b_1$), add-to-cart ($b_2$), and purchase ($b_3$) transition from exploration to conversion.

\vspace{2pt}
\noindent\textbf{Temporal User-Item Interaction Sequences.} For each user $u \in \mathcal{U}$, their interaction history is represented as a chronologically ordered sequence:
\begin{equation}
\mathcal{S}_u = [(v_1, b_1, t_1), (v_2, b_2, t_2), \ldots, (v_n, b_n, t_n)]
\end{equation}
where the triplet $(v_i, b_i, t_i)$ indicates that user $u$ interacted with item $v_i \in \mathcal{V}$ through behavior $b_i \in \mathcal{B}$ at timestamp $t_i$.

\vspace{2pt}
\noindent\textbf{Autoregressive Generation Objective.} We learn a probabilistic model $p_\theta$ that generates next $\text{k}$ interaction given history:
\begin{equation}
P(v_{n+1}, b_{n+1} | \mathcal{S}_u; \theta) = \prod_{t=1}^{T} P(x_t | x_{<t}, \mathcal{S}_u; \theta)
\end{equation}
where $\{x_t\}_{t=1}^T$ denotes the tokenized sequence and $\theta$ represents model parameters.

\begin{table*}[t]
    \centering
    \caption{Overall performance comparison on RetailRocket, Taobao, Tmall, and Insurance Dataset datasets. The sequence length $(L)$ is set to 25 for Retail Rocket, and 50 for others. The best results are highlighted in \textbf{bold}, and the second-best results are \underline{underlined}. Parameters are measured in Millions (M).}
    \vspace{-8pt}
    \label{tab:eval_retail_taobao}
    
    \footnotesize
    \setlength{\tabcolsep}{1pt}
    
    \begin{tabular}{c l cccccc c cccccc}
        \toprule
        \multirow{2}{*}{\textbf{Model Type}} & \multirow{2}{*}{\textbf{Model}} & \multicolumn{6}{c}{\textbf{RetailRocket}} & & \multicolumn{6}{c}{\textbf{Taobao}} \\
        \cmidrule(lr){3-8} \cmidrule(lr){10-15}
         & & \textbf{H@10} & \textbf{H@50} & \textbf{N@10} & \textbf{N@50} & \textbf{MRR} & \textbf{\#Param.} & & \textbf{H@10} & \textbf{H@50} & \textbf{N@10} & \textbf{N@50} & \textbf{MRR} & \textbf{\#Param.} \\
        \midrule
        
        \multirow{3}{*}{\shortstack[c]{Sequential \\ Recommendation}}
          & BERT4Rec & 0.0586 & 0.0927 & 0.0345 & 0.0420 & 0.0294 & 30.9M & & 0.0040 & 0.0079 & 0.0023 & 0.0031 & 0.0024 & 533.9M \\
          & GRU4Rec  & 0.0886 & 0.1448 & 0.0507 & 0.0632 & 0.0427 & 31.7M & & 0.0287 & \underline{0.0675} & 0.0152 & 0.0237 & 0.0139 & 534.1M \\
          & SASRec   & \underline{0.0927} & 0.1495 & 0.0550 & \underline{0.0677} & \underline{0.0471} & 30.6M & & \underline{0.0317} & 0.0655 & \underline{0.0178} & \underline{0.0251} & \underline{0.0160} & 532.9M \\ 
        \midrule

        \multirow{3}{*}{\shortstack[c]{Multi-Behavior \\ Recommendation}}
          & PBAT & 0.0830 & 0.1362 & 0.0495 & 0.0613 & 0.0426 & 32.7M & & 0.0135 & 0.0360 & 0.0069 & 0.0117 & 0.0067 & 535.9M \\
          & MBHT & 0.0809 & 0.1329 & 0.0479 & 0.0594 & 0.0411 & 30.9M & & 0.0169 & 0.0435 & 0.0086 & 0.0143 & 0.0081 & 533.6M \\
          & COPF & 0.0002 & 0.0011 & 0.0001 & 0.0003 & 0.0001 & 31.3M & & $>10$ d & $>10$ d & $>10$ d & $>10$ d & $>10$ d & 534.2M \\
        \midrule

          & HSTU$_{32}$ & 0.0795 & 0.1242 & 0.0465 & 0.0565 & 0.0393 & 30.3M & & 0.0246 & 0.0584 & 0.0131 & 0.0205 & 0.0121 & 533.0M \\
          & HSTU$_{48}$ & 0.0785 & 0.1198 & 0.0453 & 0.0545 & 0.0379 & 30.3M & & 0.0244 & 0.0579 & 0.0128 & 0.0201 & 0.0118 & 533.1M \\
          & HSTU$_{64}$ & 0.0772 & 0.1163 & 0.0447 & 0.0534 & 0.0373 & 30.4M & & 0.0254 & 0.0602 & 0.0136 & 0.0212 & 0.0126 & 533.3M \\
          & MBGen & 0.0909 & 0.1466 & 0.0518 & 0.0642 & 0.0434 & 30.5M & & 0.0176 & 0.0441 & 0.0092 & 0.0149 & 0.0088 & 533.2M \\
          & GAMER & 0.0207 & 0.0408 & 0.0120 & 0.0164 & 0.0109 & 34.4M & &
          0.0021 & 0.0074 & 0.0010 & 0.0021 & 0.0014 & 537.1M \\
          & BLADE & 0.0931 & \underline{0.1514} & 0.0547 & 0.0677 & 0.0466 & 36.8M & & 0.0232 &  0.0582 & 0.0120 & 0.0196 & 0.0111 & 538.5M\\
          \rowcolor{gray!10}
          \cellcolor{white} & \textbf{BITRec} & \textbf{0.0952} & \textbf{0.1533} & \textbf{0.0552} & \textbf{0.0681} & \textbf{0.0501} & 32.2M & & \textbf{0.0373} & \textbf{0.0765} & \textbf{0.0209} & \textbf{0.0294} & \textbf{0.0186} & 538.9M \\
          \rowcolor{gray!10}
          \cellcolor{white}\multirow{-8}{*}{\shortstack[c]{Generative \\ Recommendation}}
          & \#Improve & +2.67\% & +2.52\% & +0.28\% & +0.64\% & +6.52\% & N/A & & +17.83\% & +13.44\% & +17.55\% & +17.11\% & +16.55\% & N/A \\
        
        \midrule
        \midrule
        
        \multirow{2}{*}{\textbf{Model Type}} & \multirow{2}{*}{\textbf{Model}} & \multicolumn{6}{c}{\textbf{Tmall}} & & \multicolumn{6}{c}{\textbf{Ant Insurance Dataset}} \\
        \cmidrule(lr){3-8} \cmidrule(lr){10-15}
         & & \textbf{H@10} & \textbf{H@50} & \textbf{N@10} & \textbf{N@50} & \textbf{MRR} & \textbf{\#Param.} & & \textbf{H@10} & \textbf{H@50} & \textbf{N@10} & \textbf{N@50} & \textbf{MRR} & \textbf{\#Param.} \\
        \midrule
        
        \multirow{3}{*}{\shortstack[c]{Sequential \\ Recommendation}}
          & BERT4Rec & 0.0090 & 0.0197 & 0.0050 & 0.0073 & 0.0048 & 800.0M & & 0.0764 & 0.1267 & 0.0449 & 0.0560 & 0.0422 & 1.2M \\
          & GRU4Rec  & 0.0326 & 0.0617 & 0.0189 & 0.0253 & 0.0168 & 800.1M & & 0.0842 & 0.1466 & 0.0511 & 0.0649 & 0.0484 & 2.1M \\
          & SASRec   & \underline{0.0394} & \underline{0.0694} & \underline{0.0210} & \underline{0.0277} & \underline{0.0175} & 800.0M & & 0.0801 & 0.1427 & 0.0480 & 0.0618 & 0.0457 & 0.6M \\
        \midrule

        \multirow{3}{*}{\shortstack[c]{Multi-Behavior \\ Recommendation}}
          & PBAT & 0.0238 & 0.0482 & 0.0134 & 0.0188 & 0.0121 & 802.9M & & 0.0816 & 0.1464 & 0.0479 & 0.0622 & 0.0451 & 4.6M \\
          & MBHT & 0.0232 & 0.0467 & 0.0130 & 0.0181 & 0.0117 & 800.6M & & 0.0828 & 0.1480 & 0.0494 & 0.0639 & 0.0468 & 1.2M \\
          & COPF & $>10$ d & $>10$ d & $>10$ d & $>10$ d & $>10$ d & 806.6M & & - & - & - & - & - & 1.9M \\
        \midrule

          & HSTU$_{32}$ & 0.0266 & 0.0485 & 0.0157 & 0.0205 & 0.0140 & 800.1M & & 0.0849 & 0.1435 & 0.0523 & 0.0653 & 0.0496 & 0.6M \\
          & HSTU$_{48}$ & 0.0225 & 0.0424 & 0.0134 & 0.0177 & 0.0121 & 800.2M & & 0.0852 & 0.1436 & 0.0527 & 0.0657 & 0.0500 & 0.6M \\
          & HSTU$_{64}$ & 0.0226 & 0.0422 & 0.0136 & 0.0179 & 0.0123 & 800.4M & & \underline{0.0877} & 0.1436 & \underline{0.0550} & \underline{0.0674} & \underline{0.0521} & 0.9M \\
          & MBGen & 0.0271 & 0.0536 & 0.0153 & 0.0211 & 0.0137 & 800.3M & & 0.0813 & 0.1458 & 0.0485 & 0.0628 & 0.0461 & 0.9M \\
          & GAMER & 0.0193 & 0.0423 & 0.0105 & 0.0155 & 0.0097 & 804.1M & & 0.0829 &  0.1405 & 0.1597 & 0.0620 & 0.0424 & 4.6M \\
          & BLADE & 0.0263 & 0.0540 & 0.0147 & 0.0208 & 0.0132 & 805.6M & & 0.0869 &  \textbf{0.1502} & 0.0519 & 0.0659 & 0.0449 & 11.6M\\
          \rowcolor{gray!10}
          \cellcolor{white} & \textbf{BITRec} & \textbf{0.0407} & \textbf{0.0701} & \textbf{0.0244} & \textbf{0.0309} & \textbf{0.0215} & 805.1M & & \textbf{0.0917} & \underline{0.1501} & \textbf{0.0563} & \textbf{0.0693} & \textbf{0.0527} & 1.7M \\
          \rowcolor{gray!10}
          \cellcolor{white}\multirow{-8}{*}{\shortstack[c]{Generative \\ Recommendation}}
          & \#Improve & +3.31\% & +1.04\% & +16.09\% & +11.65\% & +22.79\% & N/A & & +4.56\% & -0.07\% & +2.36\% & +2.82\% & +1.15\% & N/A \\
        \bottomrule
    \end{tabular}
\end{table*}

\subsection{Joint Behavior-Item Embedding}

\noindent
Existing generative methods typically tokenize behaviors and items as independent tokens (e.g., [click, item A]). This interleaved design leads to sequence dilation, doubling the sequence length and making long-range dependency modeling more difficult, while also introducing semantic ambiguity between loosely coupled behavior–item pairs.

To address this issue, we construct a unified token by fusing item and behavior semantics via additive embedding. Specifically, for each interaction at position $i$ with item $v_i$ and behavior $b_i$, we compute:

 \begin{equation}
\mathbf{x}_i = \mathbf{e}_{v_i}^{\text{item}} + \mathbf{e}_{b_i}^{\text{behavior}}
\end{equation}

where $\mathbf{e}_{v_i}^{\text{item}}, \mathbf{e}_{b_i}^{\text{behavior}} \in \mathbb{R}^d$. This joint representation explicitly binds each behavior to its corresponding item at the token level, eliminating positional ambiguity and serving as the fundamental unit for subsequent hierarchical behavior aggregation, enabling the model to focus on interaction-level dependencies.

\subsection{Hierarchical Behavior Aggregation (HBA)}

Standard Transformers \cite{attention} implicitly assume uniform dependency activation, conflating shallow exploration with deep commitment. However, user behaviors exhibit significant heterogeneity in intensity \cite{NMTR,mbsr,2016wide}: shallow behaviors (e.g., clicks) reflect broad interest exploration, while deep behaviors (e.g., carting) signal conversion propensity. To disentangle these signals, we propose \textbf{Hierarchical Behavior Aggregation (HBA)}, which decomposes historical interactions into exploration and commitment pathways and adaptively weights these perspectives during prediction.

\vspace{2pt}
\noindent\textbf{Intensity-Stratified Aggregation.} Firstly, we partition the behavior set $\mathcal{B}$ into low-intensity (exploration) and high-intensity (commitment) subsets via a mapping $\phi: \mathcal{B} \rightarrow \{0, 1\}$. Let $\mathbf{H} \in \mathbb{R}^{L \times d}$ denote the input sequence representation.

We then construct two parallel aggregation channels using masked self-attention, where $i$ and $j$ index the query and key positions, respectively. The exploration representation $\mathbf{R}^{\text{low}}$ aggregates only low-intensity history:
\begin{equation}
\mathbf{M}_{ij}^{\text{low}} = \begin{cases}
0 & \text{if } \phi(b_j)=0 \text{ and } j \le i \\
-\infty & \text{otherwise}
\end{cases}
\end{equation}
\begin{equation}
\mathbf{R}^{\text{low}} = \text{Softmax}\left(\frac{\mathbf{H}\mathbf{W}_Q (\mathbf{H}\mathbf{W}_K)^\top}{\sqrt{d}} + \mathbf{M}^{\text{low}}\right) \mathbf{H}\mathbf{W}_V
\end{equation}

Simultaneously, the commitment representation $\mathbf{R}^{\text{high}}$ is computed using a complementary mask $\mathbf{M}^{\text{high}}$ where $\phi(b_j)=1$. This yields two distinct context views: $\mathbf{R}^{\text{low}}$ encodes the breadth of user interests, while $\mathbf{R}^{\text{high}}$ captures clear purchase signals.

\vspace{2pt}
\noindent\textbf{Intent-Aware Fusion via MoE.}
To enable selective dependency activation, we fuse these views based on the \textit{current} step's behavior $b_i$. We assign primary and auxiliary representations based on the query's behavior type:
\begin{equation}
(\mathbf{R}_i^{\text{self}}, \mathbf{R}_i^{\text{cross}}) = \begin{cases}
(\mathbf{R}^{\text{low}}_i, \mathbf{R}^{\text{high}}_i) & \text{if } \phi(b_i) = 0 \\
(\mathbf{R}^{\text{high}}_i, \mathbf{R}^{\text{low}}_i) & \text{if } \phi(b_i) = 1
\end{cases}
\end{equation}
where $\mathbf{R}^{\text{self}}$ aligns with the query's intensity level (intra-stratum) and $\mathbf{R}^{\text{cross}}$ provides complementary cross-stratum context.

A mixture-of-experts gating network then determines the fusion ratio:
\begin{equation}
\mathbf{g}_i = \sigma\left(\mathbf{W}_g [\mathbf{R}_i^{\text{self}} \| \mathbf{R}_i^{\text{cross}}] + \mathbf{b}_g \right)
\end{equation}
\begin{equation}
\mathbf{R}_i^{\text{fused}} = \mathbf{g}_i \odot \mathbf{R}_i^{\text{self}} + (1 - \mathbf{g}_i) \odot \mathbf{R}_i^{\text{cross}}
\end{equation}
where $\mathbf{W}_g \in \mathbb{R}^{d \times 2d}$ and $\mathbf{b}_g \in \mathbb{R}^d$ are learnable parameters. The gating vector $\mathbf{g}_i \in [0,1]^d$ performs dimension-wise mixing based on the current behavioral context.

\vspace{2pt}
\noindent\textbf{Integration.}
Finally, we transform this fused representation into an attention bias that modulates how the model attends to each historical position:
\begin{equation}
b_{ij}^{\text{HBA}} = \text{MLP}\left([\mathbf{h}_i \| \mathbf{e}_{b_j}^{\text{behavior}} \| \mathbf{e}_{v_j}^{\text{item}} \| \mathbf{R}_{i}^{\text{fused}}]\right) \cdot \mathbbm{1}(j \leq i)
\end{equation}

This hierarchical aggregation allows the model to adaptively shift focus: emphasizing browsing history when the user is exploring at position $i$, while prioritizing cart/purchase history when the user shows commitment.

\subsection{Transition Relation Encoding (TRE)}

While HBA disentangles behavioral intensities, it primarily processes interactions based on content similarity. However, user decision paths follow inherent \textit{structural dependencies}---for instance, when predicting purchase, recent add-to-cart actions should receive higher weight as direct predecessors, while early impressions contribute less despite potential semantic similarity \cite{ma2019,Xi2021,pi2019,wen2020,zhou2018}. Standard attention mechanisms lack explicit priors to enforce these transition rules.

To bridge the gap, we propose \textbf{Transition Relation Encoding (TRE)}, which injects structural inductive biases into the sequence modeling. TRE operates by learning a transition relevance function that explicitly measures the structural affinity between the historical position $j$ and the current target $i$. This is achieved through two complementary mechanisms: implicit context matching and explicit relational structure encoding.

\vspace{3pt}
\noindent\textbf{Implicit Context Matching.} First, we capture implicit dependencies by learning a specialized query-key matching mechanism. For each historical position $j$, we construct a comprehensive context vector $\mathbf{C}_j$ by integrating behavioral semantics, category intent, and statistical engagement attributes:
\begin{equation}
    \mathbf{C}_j = [\mathbf{e}_{b_j}^{\text{behavior}} \| \mathbf{e}_{c_{v_j}}^{\text{category}} \| \mathcal{H}_j]
\end{equation}
where $\mathbf{e}_{b_j}^{\text{behavior}}$ and $\mathbf{e}_{c_{v_j}}^{\text{category}}$ denote behavior and category embeddings, respectively. $\mathcal{H}_j \in \mathbb{R}^{d_h}$ encodes statistical context features, including sequence length, item/category repetition counts, and high-intensity behavior ratios, which provide dense signals regarding user engagement depth.

We then project these contexts into a latent dependency space to compute a matching score:
\begin{equation}
    \mathbf{Q}_{\text{ctx}} = \mathbf{C}\mathbf{W}_Q, \quad \mathbf{K}_{\text{ctx}} = \mathbf{C}\mathbf{W}_K, \quad
    s_{ij}^{\text{qk}} = \mathbf{Q}_{\text{ctx},i}^\top \mathbf{K}_{\text{ctx},j}
\end{equation}
This score $s_{ij}^{\text{qk}}$ represents the data-driven ``soft'' alignment between the current decision requirement and historical information availability.

\vspace{3pt}
\noindent\textbf{Explicit Relational Encoding.} To complement the implicit matching, we inject explicit domain priors regarding item consistency, behavioral logic, and temporal dynamics. We construct a \textbf{Relational Tensor} $\mathcal{R} \in \mathbb{R}^{L \times L \times d_r}$ by stacking multiple pairwise interaction maps.
\begin{itemize}[leftmargin=*]
    \item \textbf{Item-Level Consistency:} We encode whether the interaction pair targets the same item or category, along with their semantic proximity:
    \begin{equation}
        \mathbf{r}^{\text{item}}_{ij} = [\mathbbm{1}(v_i = v_j), \mathbbm{1}(c_i = c_j), \cos(\mathbf{e}_{v_i}, \mathbf{e}_{v_j})]
    \end{equation}
    
    \item \textbf{Global Behavioral Transition:} Unlike rigid heuristic rules, we employ a learnable transition matrix $\mathbf{B} \in \mathbb{R}^{|\mathcal{B}| \times |\mathcal{B}|}$ to capture global pairwise affinities (e.g., the likelihood of $\text{Click} \rightarrow \text{Purchase}$). The specific entry $B_{b_i, b_j}$ is retrieved to represent the inherent structural connection between behavior types.
    
    \item \textbf{Temporal Dynamics:} To model the decaying influence of historical interactions, we encode the time interval $\Delta t_{ij}$ using multi-scale features:
    \begin{equation}
        \mathbf{r}_{ij}^{\text{time}} = \left[ \sigma(\Delta t_{ij}/24), \log(1+\Delta t_{ij}), (1+\Delta t_{ij})^{-1} \right]
    \end{equation}
\end{itemize}

\noindent These explicit relations are concatenated and fused via a relational MLP:
\begin{equation}
    s_{ij}^{\text{rel}} = \text{MLP}_{\text{rel}}([\mathbf{r}^{\text{item}}_{ij} \| B_{b_i, b_j} \| \mathbf{r}_{ij}^{\text{time}}])
\end{equation}

\vspace{2pt}
\noindent\textbf{Fusion and Bias Injection.} Finally, we combine the data-driven matching $s_{ij}^{\text{qk}}$ and structured relational score $s_{ij}^{\text{rel}}$ to derive the final transition weight:
\begin{equation}
    w_{ij}^{\text{trans}} = \sigma\left(\frac{\alpha_{\text{qk}}}{\tau} s_{ij}^{\text{qk}} + \alpha_{\text{rel}} \cdot s_{ij}^{\text{rel}}\right)
\end{equation}
\begin{equation}
    b_{ij}^{\text{TRE}} = \log(w_{ij}^{\text{trans}} + \epsilon) \cdot \mathbbm{1}(j \leq i)
\end{equation}
where $\alpha_{\text{qk}}, \alpha_{\text{rel}}$ are learnable gates balancing implicit and explicit signals, and $\tau$ is a temperature scaling factor. The resulting term $b_{ij}^{\text{TRE}}$ serves as an additive bias in the attention mechanism, strictly enforcing structural validity in the generative process.

\begin{figure*}[!t]
    \centering
    
    \begin{subfigure}{\textwidth}
        \centering
        \includegraphics[width=\textwidth]{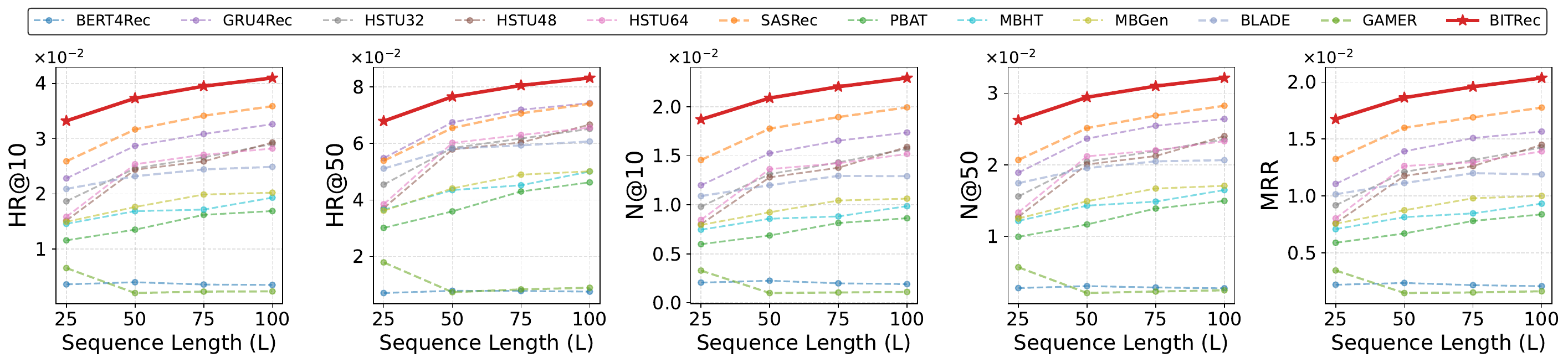}
         \vspace{-0.6cm}   
        \caption{Performance comparison cross different sequence lengths on the Taobao dataset.}
        \label{fig:ablation_taobao}
    \end{subfigure}
    
    
    \begin{subfigure}{\textwidth}
        \centering
        \includegraphics[width=\textwidth]{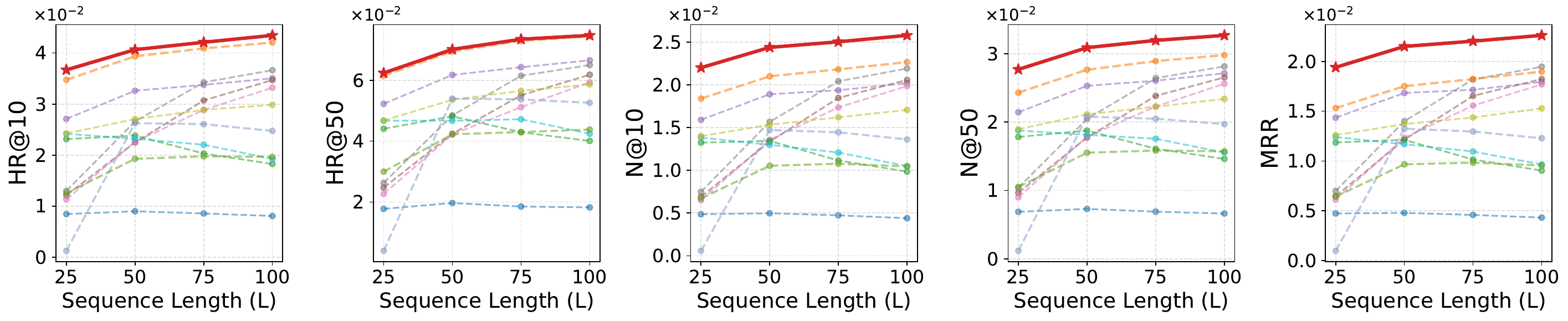}
        \vspace{-0.6cm}
        \caption{Performance comparison cross different sequence lengths on the Tmall dataset.}
        \label{fig:ablation_tmall}
    \end{subfigure}
    
    
    \vspace{-10pt}
    \caption{Performance comparison across different sequence lengths ($L$) on (a) Taobao, (b) Tmall. 
    \textbf{BITRec (red)} consistently outperforms all baselines and maintains stable improvements as sequence length increases, 
    while some baselines like HSTU (cyan) show performance degradation on longer sequences. 
    This validates BITRec's ability to effectively leverage long-range multi-behavior dependencies through joint embedding and explicit structure modeling.}
    \vspace{-15pt}
    \label{fig:ablation_study_sequence_length}
\end{figure*}

\subsection{Attention Integration}

We integrate intensity biases from HBA and transition biases from TRE into standard self-attention. The final attention score combines three components:
\begin{equation}
s_{ij} = \frac{\mathbf{q}_i^\top \mathbf{k}_j}{\sqrt{d_k}} + \beta \cdot b_{ij}^{\text{HBA}} + \gamma \cdot b_{ij}^{\text{TRE}}
\end{equation}
where $\beta, \gamma$ are learnable weights balancing content similarity, behavioral intensity, and transition structure. Attention weights are computed as:
\begin{equation}
\alpha_{ij} = \frac{\exp(s_{ij})}{\sum_{k=1}^{n} \exp(s_{ik})}, \quad \mathbf{h}'_i = \sum_{j=1}^{n} \alpha_{ij} \mathbf{v}_j
\end{equation}
This enables selective attention where intensity bias emphasizes positions aligned with query intensity and transition bias prioritizes relevant funnel positions.

\input{section/exp}

\section{Conclusion}

In this paper, we propose BITRec, a generative framework that introduces structured behavioral modeling into multi-behavior recommendation through selective dependency activation. BITRec addresses behavioral intensity heterogeneity and missing transition dependencies through Hierarchical Behavior Aggregation (HBA) and Transition Relation Encoding (TRE), respectively. Experiments on four large-scale datasets (RetailRocket, Taobao, Tmall, Insurance Dataset) achieve consistent improvements of 15-23\% across metrics, with particularly strong gains on ranking metrics: 17.55\% NDCG@10 and 16.55\% MRR on Taobao, 22.79\% MRR on Tmall. Ablations confirm both HBA and TRE contribute 2-7\% individually, validating that selective dependency modeling is essential for multi-behavior recommendation.

\bibliographystyle{named}
\bibliography{ijcai26}


\end{document}


\maketitle

\section{Additional Experimental Results}

\subsection{Implementation Details}

Table~\ref{tab:hyperparameters} summarizes the hyperparameters used across all datasets. All experiments are conducted on 8 NVIDIA A100 GPUs with a total batch size of 2048 (256 per GPU). We use the Adam \cite{adam} optimizer with weight decay of 0.01 and employ a cosine learning rate scheduler with warmup for the first 10\% of training steps.

For training, we adopt the sampled softmax loss with 128 negative samples per positive instance, which provides an efficient approximation to full softmax while maintaining training stability. The embedding dimension is set to 128 for all models to ensure fair comparison. For HBA, we partition behaviors into exploration (exposure, click) and commitment (search, add-to-cart, consult, purchase) groups based on behavioral intensity semantics. For TRE, we use 2 attention heads with learnable behavior relation matrices of size $|\mathcal{B}| \times |\mathcal{B}|$ where $|\mathcal{B}|$ is the number of behavior types.

\begin{table}[t]
\centering
\caption{Hyperparameter settings across datasets.}
\label{tab:hyperparameters}
\footnotesize
\setlength{\tabcolsep}{4pt}
\begin{tabular}{l cccc}
\toprule
\textbf{Hyperparameter} & \textbf{RetailRocket} & \textbf{Taobao} & \textbf{Tmall} & \textbf{Insurance} \\
\midrule
Learning Rate & 2e-4 & 3e-4 & 2e-4 & 2e-4 \\
Epochs & 80 & 50 & 80 & 50 \\
Batch Size & \multicolumn{4}{c}{2048 (256 $\times$ 8 GPUs)} \\
Negative Samples & \multicolumn{4}{c}{128} \\
Embedding Dim & \multicolumn{4}{c}{128} \\
LR Scheduler & \multicolumn{4}{c}{Cosine with 10\% warmup} \\
\# Attention Heads & \multicolumn{4}{c}{2} \\
\bottomrule
\end{tabular}
\end{table}

\subsection{Baseline Implementation Details}

All baselines are implemented using their official codebases or reproduced following their original papers. For fair comparison, we unify the embedding dimension to 128 and use the same negative sampling strategy (128 negatives).

All baselines use 4 Transformer layers \cite{attention} with 2 attention heads and embedding dimension of 128 for fair comparison. Model-specific settings: \textbf{BERT4Rec} uses mask ratio 0.2; \textbf{GRU4Rec} uses 4 GRU layers with dropout 0.2; \textbf{HSTU} is evaluated with 32, 48, and 64 attention heads; \textbf{PBAT} adds personalized behavior-aware attention; \textbf{MBHT} incorporates hypergraph convolution; \textbf{MBGen} uses position-routed sparse architecture.

\subsection{Fair Comparison and Evaluation Protocol} All models use similar capacity (embedding dimension 128, comparable layers/heads) ensuring fair comparison. Our absolute metrics are notably lower than some prior work due to different evaluation protocols. Most existing methods (MBHT, PBAT, MB-GMN, NMTR) adopt negative sampling, pairing each positive with 100 negatives (e.g., MBHT achieves 25.7\% NDCG@5 on Taobao ranking among 101 candidates). Following HSTU \cite{hstu}, we adopt full-item ranking across complete catalogs (RetailRocket: $\sim$235K items, Taobao: $\sim$4.16M items). While yielding lower absolute values, this better reflects real-world difficulty. BITRec's relative improvements (19.69\% Taobao HR@10, 22.17\% Tmall MRR) validate our method under this rigorous protocol.

\subsection{Robustness under Behavior Masking}
\begin{figure}[!t]
\centering
\includegraphics[width=\columnwidth]{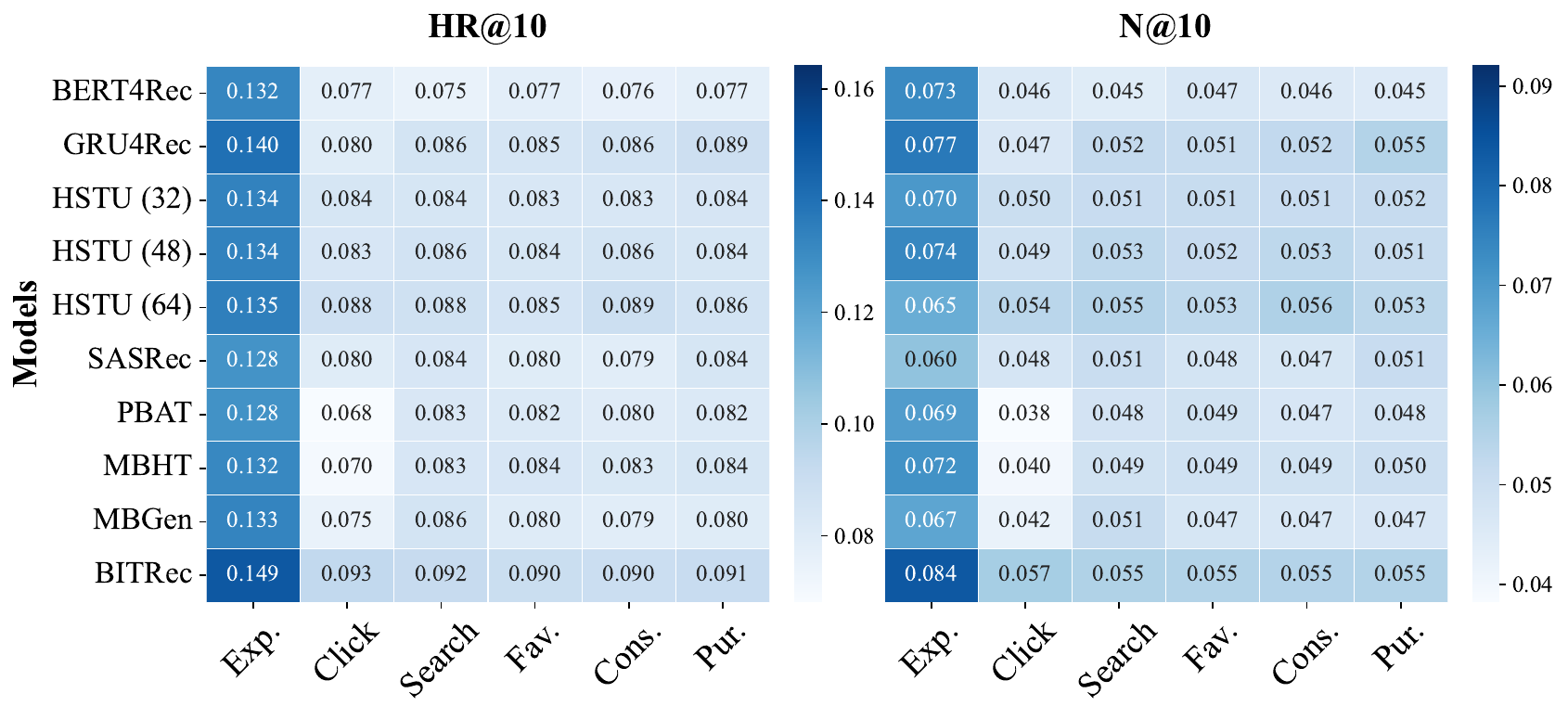}
\vspace{-20pt}
\caption{Model robustness under behavior masking on Insurance Dataset. Each column shows performance when a specific behavior is masked during training. Lighter colors indicate better performance. BITRec (bottom row) consistently outperforms baselines across all masking scenarios, demonstrating superior ability to handle incomplete behavioral data through selective dependency activation.}
\vspace{-10pt}
\label{fig:behavior_robustness}
\end{figure}

We evaluate model robustness by masking each behavior type during training and testing on complete sequences. Figure~\ref{fig:behavior_robustness} shows results on Insurance Dataset, which contains 6 behavior types. BITRec maintains the best performance across all masking scenarios, as shown by the lightest colors in the bottom row. When deep behaviors (Consult, Purchase) are masked, baseline models suffer significant performance drops while BITRec remains relatively stable, validating HBA's ability to leverage shallow behaviors effectively when commitment signals are unavailable. Moreover, BITRec shows uniform degradation patterns across different maskings, indicating that TRE successfully captures structured dependencies rather than over-relying on specific behaviors. This robustness is crucial for real-world deployment, where complete behavioral data is often unavailable.

\begin{table*}[t]
    \centering
    \caption{Performance comparison on the Taobao dataset across different sequence lengths ($L=\{25, 75, 100\}$). All metrics are reported in absolute values (original percentage values divided by 100). The best results are highlighted in \textbf{bold}, and the second-best results are \underline{underlined}. The row ``\#Improve'' indicates the relative improvement of BITRec over the best baseline.}
    \label{tab:eval_taobao_wide_improve}
    \vspace{-5pt}
    
    \footnotesize
    \setlength{\tabcolsep}{1.5pt}
    
    \begin{tabular}{l ccccc c ccccc c ccccc}
        \toprule
        \multirow{2}{*}{\textbf{Model}} & \multicolumn{5}{c}{\textbf{Taobao ($L=25$)}} & & \multicolumn{5}{c}{\textbf{Taobao ($L=75$)}} & & \multicolumn{5}{c}{\textbf{Taobao ($L=100$)}} \\
        \cmidrule(lr){2-6} \cmidrule(lr){8-12} \cmidrule(lr){14-18}
         & \textbf{H@10} & \textbf{H@50} & \textbf{N@10} & \textbf{N@50} & \textbf{MRR} & & \textbf{H@10} & \textbf{H@50} & \textbf{N@10} & \textbf{N@50} & \textbf{MRR} & & \textbf{H@10} & \textbf{H@50} & \textbf{N@10} & \textbf{N@50} & \textbf{MRR} \\
        \midrule
        
        BERT4Rec    & 0.0036 & 0.0071 & 0.0021 & 0.0028 & 0.0022 & & 0.0036 & 0.0078 & 0.0020 & 0.0029 & 0.0022 & & 0.0035 & 0.0076 & 0.0019 & 0.0028 & 0.0021 \\
        GRU4Rec     & 0.0228 & \underline{0.0548} & 0.0120 & 0.0189 & 0.0111 & & 0.0309 & \underline{0.0719} & 0.0165 & 0.0255 & 0.0151 & & 0.0326 & \underline{0.0743} & 0.0174 & 0.0264 & 0.0157 \\
        HSTU$_{32}$ & 0.0187 & 0.0454 & 0.0098 & 0.0156 & 0.0092 & & 0.0266 & 0.0617 & 0.0143 & 0.0219 & 0.0132 & & 0.0291 & 0.0652 & 0.0157 & 0.0236 & 0.0143 \\
        HSTU$_{48}$ & 0.0152 & 0.0371 & 0.0080 & 0.0128 & 0.0076 & & 0.0259 & 0.0603 & 0.0138 & 0.0212 & 0.0126 & & 0.0293 & 0.0667 & 0.0159 & 0.0240 & 0.0145 \\
        HSTU$_{64}$ & 0.0159 & 0.0385 & 0.0085 & 0.0134 & 0.0080 & & 0.0271 & 0.0630 & 0.0142 & 0.0220 & 0.0130 & & 0.0282 & 0.0656 & 0.0152 & 0.0233 & 0.0139 \\
        SASRec      & \underline{0.0259} & 0.0539 & \underline{0.0146} & \underline{0.0207} & \underline{0.0133} & & \underline{0.0342} & 0.0706 & \underline{0.0189} & \underline{0.0269} & \underline{0.0169} & & \underline{0.0359} & 0.0741 & \underline{0.0199} & \underline{0.0283} & \underline{0.0178} \\
        PBAT        & 0.0116 & 0.0301 & 0.0060 & 0.0100 & 0.0059 & & 0.0162 & 0.0430 & 0.0081 & 0.0139 & 0.0078 & & 0.0169 & 0.0462 & 0.0086 & 0.0150 & 0.0084 \\
        MBHT        & 0.0146 & 0.0365 & 0.0075 & 0.0122 & 0.0071 & & 0.0172 & 0.0452 & 0.0088 & 0.0149 & 0.0085 & & 0.0193 & 0.0501 & 0.0099 & 0.0165 & 0.0093 \\
        MBGen       & 0.0149 & 0.0362 & 0.0079 & 0.0125 & 0.0076 & & 0.0199 & 0.0490 & 0.0104 & 0.0167 & 0.0098 & & 0.0202 & 0.0501 & 0.0106 & 0.0171 & 0.0100 \\
        BLADE       & 0.0209 & 0.0511 & 0.0109 & 0.0175 & 0.0102 & & 0.0244 & 0.0593 & 0.0129 & 0.0205 & 0.0120 & & 0.0249 & 0.0607 & 0.0129 & 0.0207 & 0.0119 \\
        GAMER       & 0.0066 & 0.0179 & 0.0033 & 0.0057 & 0.0035 & & 0.0023 & 0.0084 & 0.0011 & 0.0023 & 0.0015 & & 0.0024 & 0.0089 & 0.0011 & 0.0025 & 0.0016 \\
        
        \midrule
        \rowcolor{gray!10} 
        \textbf{BITRec} & \textbf{0.0332} & \textbf{0.0679} & \textbf{0.0187} & \textbf{0.0262} & \textbf{0.0167} & & \textbf{0.0395} & \textbf{0.0805} & \textbf{0.0220} & \textbf{0.0310} & \textbf{0.0196} & & \textbf{0.0410} & \textbf{0.0832} & \textbf{0.0229} & \textbf{0.0321} & \textbf{0.0204} \\
        
        \rowcolor{gray!10}
        \#Improve & +28.3\% & +23.9\% & +28.4\% & +26.9\% & +26.4\% & 
                  & +15.7\% & +12.0\% & +16.3\% & +15.3\% & +15.9\% & 
                  & +14.2\% & +12.0\% & +15.0\% & +13.7\% & +14.6\% \\
        
        \bottomrule
    \end{tabular}
\end{table*}

\begin{table*}[t]
    \centering
    \caption{Performance comparison on the Tmall dataset across different sequence lengths ($L=\{25, 75, 100\}$). All metrics are reported in absolute values (original percentage values divided by 100). The best results are highlighted in \textbf{bold}, and the second-best results are \underline{underlined}. The row ``\#Improve'' indicates the relative improvement of BITRec over the best baseline.}
    \label{tab:eval_tmall_wide_improve}
    \vspace{-5pt}
    
    \footnotesize
    \setlength{\tabcolsep}{2pt}
    
    \begin{tabular}{l ccccc c ccccc c ccccc}
        \toprule
        \multirow{2}{*}{\textbf{Model}} & \multicolumn{5}{c}{\textbf{Tmall ($L=25$)}} & & \multicolumn{5}{c}{\textbf{Tmall ($L=75$)}} & & \multicolumn{5}{c}{\textbf{Tmall ($L=100$)}} \\
        \cmidrule(lr){2-6} \cmidrule(lr){8-12} \cmidrule(lr){14-18}
         & \textbf{H@10} & \textbf{H@50} & \textbf{N@10} & \textbf{N@50} & \textbf{MRR} & & \textbf{H@10} & \textbf{H@50} & \textbf{N@10} & \textbf{N@50} & \textbf{MRR} & & \textbf{H@10} & \textbf{H@50} & \textbf{N@10} & \textbf{N@50} & \textbf{MRR} \\
        \midrule
        
        BERT4Rec    & 0.0085 & 0.0178 & 0.0049 & 0.0069 & 0.0047 & & 0.0086 & 0.0185 & 0.0047 & 0.0069 & 0.0046 & & 0.0081 & 0.0183 & 0.0044 & 0.0066 & 0.0043 \\
        GRU4Rec     & 0.0271 & 0.0523 & 0.0159 & 0.0214 & 0.0144 & & 0.0338 & 0.0643 & 0.0194 & 0.0261 & 0.0172 & & 0.0350 & 0.0666 & 0.0202 & 0.0271 & 0.0179 \\
        HSTU$_{32}$ & 0.0130 & 0.0263 & 0.0075 & 0.0104 & 0.0070 & & 0.0343 & 0.0615 & 0.0204 & 0.0264 & 0.0182 & & 0.0366 & 0.0650 & 0.0219 & 0.0281 & \underline{0.0195} \\
        HSTU$_{48}$ & 0.0121 & 0.0247 & 0.0069 & 0.0097 & 0.0065 & & 0.0307 & 0.0550 & 0.0185 & 0.0238 & 0.0165 & & 0.0347 & 0.0618 & 0.0206 & 0.0265 & 0.0182 \\
        HSTU$_{64}$ & 0.0113 & 0.0227 & 0.0065 & 0.0090 & 0.0061 & & 0.0289 & 0.0512 & 0.0174 & 0.0223 & 0.0156 & & 0.0332 & 0.0593 & 0.0199 & 0.0256 & 0.0177 \\
        SASRec      & \underline{0.0348} & \underline{0.0615} & \underline{0.0184} & \underline{0.0243} & \underline{0.0153} & & \underline{0.0409} & \underline{0.0729} & \underline{0.0218} & \underline{0.0289} & \underline{0.0182} & & \underline{0.0421} & \underline{0.0744} & \underline{0.0227} & \underline{0.0298} & 0.0190 \\
        PBAT        & 0.0231 & 0.0441 & 0.0132 & 0.0178 & 0.0119 & & 0.0203 & 0.0429 & 0.0112 & 0.0161 & 0.0101 & & 0.0183 & 0.0401 & 0.0098 & 0.0146 & 0.0090 \\
        MBHT        & 0.0242 & 0.0467 & 0.0138 & 0.0188 & 0.0124 & & 0.0220 & 0.0472 & 0.0121 & 0.0176 & 0.0110 & & 0.0192 & 0.0424 & 0.0105 & 0.0155 & 0.0096 \\
        MBGen       & 0.0242 & 0.0468 & 0.0140 & 0.0189 & 0.0126 & & 0.0289 & 0.0565 & 0.0162 & 0.0222 & 0.0144 & & 0.0299 & 0.0586 & 0.0171 & 0.0234 & 0.0153 \\
        BLADE       & 0.0013 & 0.0040 & 0.0006 & 0.0012 & 0.0010 & & 0.0261 & 0.0537 & 0.0144 & 0.0205 & 0.0130 & & 0.0247 & 0.0526 & 0.0136 & 0.0197 & 0.0123 \\
        GAMER       & 0.0125 & 0.0300 & 0.0067 & 0.0105 & 0.0064 & & 0.0198 & 0.0430 & 0.0108 & 0.0158 & 0.0098 & & 0.0196 & 0.0438 & 0.0105 & 0.0157 & 0.0096 \\
        
        \midrule
        \rowcolor{gray!10} 
        \textbf{BITRec} & \textbf{0.0367} & \textbf{0.0624} & \textbf{0.0220} & \textbf{0.0277} & \textbf{0.0194} & & \textbf{0.0421} & \textbf{0.0735} & \textbf{0.0250} & \textbf{0.0320} & \textbf{0.0220} & & \textbf{0.0435} & \textbf{0.0747} & \textbf{0.0258} & \textbf{0.0327} & \textbf{0.0226} \\
        
        \rowcolor{gray!10}
        \#Improve & +5.5\% & +1.5\% & +19.6\% & +14.0\% & +26.8\% & 
                  & +2.9\% & +0.8\% & +14.7\% & +10.7\% & +20.9\% & 
                  & +3.3\% & +0.4\% & +13.7\% & +9.7\% & +15.9\% \\
        
        \bottomrule
    \end{tabular}
\end{table*}

\begin{table*}[!t]
    \centering
    \caption{Ablation study of different model components (HBA, TRE) on the Taobao dataset across different sequence lengths. The best results are highlighted in \textbf{bold}, and the second-best results are \underline{underlined}. The row ``\#Improve'' indicates the relative improvement of BITRec over the best baseline variant.}
    \vspace{-10pt}
    \label{tab:ablation_study_taobao_all}

    \footnotesize
    \setlength{\tabcolsep}{3pt}

    \begin{tabular}{l cc ccccc ccccc}
        \toprule
        \multirow{2}{*}{\textbf{Model Variant}} & \multicolumn{2}{c}{\textbf{Components}} & \multicolumn{5}{c}{\textbf{Taobao ($L=25$)}} & \multicolumn{5}{c}{\textbf{Taobao ($L=50$)}} \\
        \cmidrule(lr){2-3} \cmidrule(lr){4-8} \cmidrule(lr){9-13}
         & HBA & TRE
         & \textbf{H@10} & \textbf{H@50} & \textbf{N@10} & \textbf{N@50} & \textbf{MRR}
         & \textbf{H@10} & \textbf{H@50} & \textbf{N@10} & \textbf{N@50} & \textbf{MRR} \\
        \midrule
        
        w/o HBA
        & $\times$ & \checkmark 
        & \underline{0.0325} & 0.0661 & 0.0182 & 0.0256 & 0.0163
        & 0.0367 & 0.0756 & \underline{0.0204} & 0.0289 & 0.0181 \\
        
        w/o TRE
        & \checkmark & $\times$ 
        & 0.0324 & \underline{0.0661} & \underline{0.0183} & \underline{0.0257} & \underline{0.0165}
        & \underline{0.0368} & \underline{0.0758} & \underline{0.0204} & \underline{0.0289} & \underline{0.0182} \\

        \textbf{Full BITRec} 
        & \checkmark & \checkmark 
        & \textbf{0.0332} & \textbf{0.0679} & \textbf{0.0187} & \textbf{0.0262} & \textbf{0.0167}
        & \textbf{0.0373} & \textbf{0.0765} & \textbf{0.0209} & \textbf{0.0294} & \textbf{0.0186} \\
        
        \cmidrule(lr){2-13} 
        \rowcolor{gray!10}
        \#Improve & ~ & ~ 
        & +2.31\% & +2.67\% & +2.07\% & +2.11\% & +1.56\%
        & +1.32\% & +1.03\% & +2.22\% & +1.75\% & +2.46\% \\
        \bottomrule
    \end{tabular}
    

        
        

        
\end{table*}

\begin{table*}[!t]
    \centering
    \caption{Ablation study of different model components (HBA, TRE) on the Tmall dataset across different sequence lengths. The best results are highlighted in \textbf{bold}, and the second-best results are \underline{underlined}. The row ``\#Improve'' indicates the relative improvement of BITRec over the best baseline variant.}
    \vspace{-10pt}
    \label{tab:ablation_study_tmall_all}

    \footnotesize
    \setlength{\tabcolsep}{3pt}

    \begin{tabular}{l cc ccccc ccccc}
        \toprule
        \multirow{2}{*}{\textbf{Model Variant}} & \multicolumn{2}{c}{\textbf{Components}} & \multicolumn{5}{c}{\textbf{Tmall ($L=25$)}} & \multicolumn{5}{c}{\textbf{Tmall ($L=50$)}} \\
        \cmidrule(lr){2-3} \cmidrule(lr){4-8} \cmidrule(lr){9-13}
         & HBA & TRE
         & \textbf{H@10} & \textbf{H@50} & \textbf{N@10} & \textbf{N@50} & \textbf{MRR}
         & \textbf{H@10} & \textbf{H@50} & \textbf{N@10} & \textbf{N@50} & \textbf{MRR} \\
        \midrule
        
        w/o HBA
        & $\times$ & \checkmark 
        & 0.0360 & 0.0620 & 0.0215 & 0.0273 & 0.0190
        & 0.0402 & \textbf{0.0702} & 0.0239 & 0.0305 & 0.0210 \\
        
        w/o TRE
        & \checkmark & $\times$ 
        & \underline{0.0363} & \underline{0.0620} & \underline{0.0217} & \underline{0.0274} & \underline{0.0191}
        & \underline{0.0403} & 0.0697 & \underline{0.0242} & \underline{0.0306} & \underline{0.0213} \\

        \textbf{Full BITRec} 
        & \checkmark & \checkmark 
        & \textbf{0.0367} & \textbf{0.0624} & \textbf{0.0220} & \textbf{0.0277} & \textbf{0.0194}
        & \textbf{0.0407} & \underline{0.0701} & \textbf{0.0244} & \textbf{0.0309} & \textbf{0.0215} \\
        
        \cmidrule(lr){2-13} 
        \rowcolor{gray!10}
        \#Improve & ~ & ~ 
        & +1.12\% & +0.55\% & +1.30\% & +1.10\% & +1.36\%
        & +0.80\% & -0.04\% & +0.75\% & +0.80\% & +0.75\% \\
        \bottomrule
    \end{tabular}
    
    \par   \smallskip 

        
        

        
\end{table*}

\subsection{Performance Across Different Lengths}

To evaluate BITRec's scalability with varying sequence lengths, we conduct experiments on Taobao and Tmall datasets with $L \in \{25, 50, 75, 100\}$.

Table~\ref{tab:eval_taobao_wide_improve} shows performance on Taobao. BITRec consistently outperforms all baselines across all sequence lengths, achieving 12-28\% improvements. Gains are most pronounced at shorter sequences and remain substantial at longer sequences ($L=100$: 14-15\%).

Table~\ref{tab:eval_tmall_wide_improve} presents results on Tmall. BITRec achieves the best performance across all settings, with particularly strong gains on ranking metrics (15.9-26.8\% MRR improvements). The consistent superiority validates that selective dependency activation effectively captures behavioral patterns regardless of sequence length.

\subsection{Ablation Study Across Sequence Lengths}

Tables~\ref{tab:ablation_study_taobao_all} and \ref{tab:ablation_study_tmall_all} present ablation results on Taobao and Tmall with $L \in \{25, 50\}$.

Both HBA and TRE consistently contribute to improvements regardless of sequence length. On Taobao, gains are slightly larger at shorter sequences ($L=25$: 1.6-2.7\%) compared to longer sequences ($L=50$: 1.0-2.5\%), suggesting that explicit behavioral modeling is particularly valuable when limited context is available. On Tmall, contributions are more modest (0.5-1.4\%) but remain consistently positive across both sequence lengths.

\bibliographystyle{named}
\bibliography{ijcai26}

%% file: section/exp.tex
\section{Experiments}

\subsection{Settings}
\textbf{Datasets.} We conduct experiments on four large-scale e-commerce benchmarks: \textbf{RetailRocket} \cite{Retailrocket}, \textbf{Taobao} \cite{taobao}, \textbf{Tmall} \cite{Tmall}, and \textbf{Ant Insurance Dataset}\footnote{A dataset collecting from Alipay Insurance Platform contains six types of user behavior: page view, click, consult, search, add-to-chart and purchase.}. Table~\ref{tab:dataset_statistics_transposed} summarizes the statistics. 

\vspace{2pt}
\noindent\textbf{Baselines.} We compare BITRec with three categories of state-of-the-art methods: (1) \textbf{Sequential Models}: GRU4Rec \cite{gru4rec}, SASRec \cite{SASRec}, BERT4Rec \cite{BERT4Rec}; (2) \textbf{Multi-Behavior Models}: MBHT \cite{mbht}, PBAT \cite{PBAT}, COPF \cite{copf}; (3) \textbf{Generative Methods}: the hierarchical model HSTU \cite{hstu}, MBGen \cite{MBgen}, which represents the current generative SOTA.

\vspace{2pt}
\noindent
More details including hyperparameters, training configurations, and baseline settings are provided in Appendix.

\subsection{Comparison With SOTA Methods}

Table~\ref{tab:eval_retail_taobao} presents performance comparisons across four datasets, we highlight three key findings.

\vspace{2pt}
\noindent\textbf{BITRec achieves consistent improvements across all settings.} BITRec outperforms all baselines on every dataset and metric combination. The improvements are particularly pronounced on Taobao, where BITRec surpasses the strongest baseline SASRec by 17.83\% on HR@10, 17.55\% on NDCG@10, and 16.55\% on MRR. This consistent superiority validates our core hypothesis: explicitly modeling behavior intensity heterogeneity (via HBA) and transition dependencies (via TRE) provides complementary benefits for multi-behavior recommendation.

\vspace{2pt}
\noindent\textbf{Ranking quality improves more than recall.} BITRec exhibits significantly higher improvements on ranking metrics (MRR, NDCG) compared to hit-based metrics (HR). For instance, on Tmall, BITRec boosts MRR by 22.79\% compared to a 3.31\% gain in HR@10 against the best baseline. This disparity reveals BITRec's core advantage: strictly distinguishing "commitment" from "exploration" via HBA enables the model to prioritize high-confidence conversion signals near the top of the recommendation list, rather than merely recalling relevant items vaguely.

\vspace{2pt}
\noindent\textbf{Robustness in Complex Dependency Scenarios.} The performance gap widens significantly on datasets with longer sequences and complex dependencies (Taobao, Tmall) compared to simpler ones (RetailRocket). While baseline improvements on RetailRocket are marginal (0.28\% NDCG), Taobao gains a 17.55\% surge. This validates the efficacy of TRE: in complex sequences where standard attention suffers from noise and signal dilution, TRE's structural transition matrix explicitly guides the model to attend to causal predecessors, filtering out irrelevant intermediate noises. We note that COPF fails to scale to Taobao and Tmall (training exceeds 10 days) and produces no results on Insurance Dataset, underscoring BITRec's efficiency advantage.

\begin{table*}[!t]
    \centering
    \caption{Ablation study on Hierarchical Behavior Aggregation (HBA) across RetailRocket, Taobao, Tmall, and Insurance Dataset datasets. The metrics reported are HR@10, NDCG@10 (N@10), and MRR. ``w/o'' denotes removing a specific component from the full BITRec model. Best results are bolded, and second-best results are underlined.}
    \vspace{-8pt}
    \label{tab:ablation_hba_compact}
    
    \footnotesize
    \setlength{\tabcolsep}{1.8pt}
    
    \begin{tabular}{l ccc c ccc c ccc c ccc}
        \toprule
        \multirow{2}{*}{\textbf{Model Variant}} & \multicolumn{3}{c}{\textbf{RetailRocket}} & & \multicolumn{3}{c}{\textbf{Taobao}} & & \multicolumn{3}{c}{\textbf{Tmall}} & & \multicolumn{3}{c}{\textbf{Insurance Dataset}} \\
        \cmidrule(lr){2-4} \cmidrule(lr){6-8} \cmidrule(lr){10-12} \cmidrule(lr){14-16}
         & \textbf{HR@10} & \textbf{N@10} & \textbf{MRR} & & \textbf{HR@10} & \textbf{N@10} & \textbf{MRR} & & \textbf{HR@10} & \textbf{N@10} & \textbf{MRR} & & \textbf{HR@10} & \textbf{N@10} & \textbf{MRR} \\
        \midrule
        
        w/o HBA & 0.0923 & 0.0542 & 0.0498 & & 0.0367 & 0.0204 & 0.0181 & & 0.0397 & 0.0235 & 0.0207 & & 0.0867 & 0.0529 & 0.0497 \\
        w/o Intensity Split & 0.0924 & 0.0543 & 0.0498 & & 0.0369 & 0.0206 & 0.0185 & & 0.0399 & 0.0236 & 0.0207 & & 0.0871 & 0.0527 & 0.0496 \\
            Purchase--Only High & \underline{0.0927} & \underline{0.0545} & \underline{0.0499} & & \underline{0.0372} & \underline{0.0208} & \underline{0.0186} & & \underline{0.0403} & \underline{0.0239} & \underline{0.0210} & & \underline{0.0882} & \underline{0.0536} & \underline{0.0503} \\
        
        \midrule
        
        \textbf{Full BITRec} & \textbf{0.0952} & \textbf{0.0552} & \textbf{0.0501} & & \textbf{0.0373} & \textbf{0.0209} & \textbf{0.0186} & & \textbf{0.0407} & \textbf{0.0244} & \textbf{0.0215} & & \textbf{0.0917} & \textbf{0.0563} & \textbf{0.0527} \\
        
        \bottomrule
    \vspace{-15pt}
    \end{tabular}
\end{table*}

\begin{table*}[t]
    \centering
    \caption{Ablation study on Transition Relation Encoding (TRE) components. We evaluate the impact of different design choices across RetailRocket, Taobao, Tmall, and Insurance Dataset datasets. The metrics reported are HR@10, NDCG@10 (N@10), and MRR. ``w/o'' denotes removing a specific component from the full BITRec model.}
    \vspace{-8pt}
    \label{tab:ablation_tre}
    
    \footnotesize
    \setlength{\tabcolsep}{3.0pt}
    
    \begin{tabular}{l ccc c ccc c ccc c ccc}
        \toprule
        \multirow{2}{*}{\textbf{Model Variant}} 
        & \multicolumn{3}{c}{\textbf{RetailRocket}} & 
        & \multicolumn{3}{c}{\textbf{Taobao}} & 
        & \multicolumn{3}{c}{\textbf{Tmall}} & 
        & \multicolumn{3}{c}{\textbf{Insurance Dataset}} \\
        \cmidrule(lr){2-4} \cmidrule(lr){6-8} \cmidrule(lr){10-12} \cmidrule(lr){14-16}
         & \textbf{H@10} & \textbf{N@10} & \textbf{MRR} &
         & \textbf{H@10} & \textbf{N@10} & \textbf{MRR} &
         & \textbf{H@10} & \textbf{N@10} & \textbf{MRR} &
         & \textbf{H@10} & \textbf{N@10} & \textbf{MRR} \\
        \midrule
        
        w/o TRE & 0.0912 & 0.0523 & 0.0476 &
                & 0.0358 & 0.0199 & 0.0179 &
                & 0.0396 & 0.0233 & 0.0205 &
                & 0.0866 & 0.0529 & 0.0499 \\
        
        w/o Global Behavioral Transition & 0.0921 & 0.0540 & 0.0495 &
                            & 0.0366 & 0.0206 & 0.0184 &
                            & 0.0400 & \underline{0.0240} & \underline{0.0213} &
                            & 0.0899 & \underline{0.0553} & \underline{0.0520} \\
        
        w/o Temporal Dynamics & 0.0917 & 0.0537 & 0.0492 &
                              & 0.0360 & 0.0202 & 0.0181 &
                              & 0.0398 & 0.0239 & 0.0211 &
                              & 0.0901 & 0.0552 & 0.0518 \\
        
        w/o Item-Level Consistency & 0.0923 & \underline{0.0546} & \underline{0.0501} &
                            & \underline{0.0373} & \underline{0.0209} & \underline{0.0186} &
                            & 0.0401 & 0.0239 & 0.0210 &
                            & \underline{0.0905} & 0.0549 & 0.0512 \\
        
        w/o Implicit Context Matching & \underline{0.0925} & 0.0545 & 0.0500 &
                         & 0.0367 & 0.0206 & 0.0185 &
                         & \underline{0.0402} & 0.0239 & 0.0210 &
                         & 0.0901 & 0.0548 & 0.0513 \\
        
        \midrule
        
        \textbf{Full BITRec} & \textbf{0.0952} & \textbf{0.0552} & \textbf{0.0501} &
                             & \textbf{0.0373} & \textbf{0.0209} & \textbf{0.0186} &
                             & \textbf{0.0407} & \textbf{0.0244} & \textbf{0.0215} &
                             & \textbf{0.0917} & \textbf{0.0563} & \textbf{0.0527} \\
        
        \bottomrule
        \vspace{-20pt}
    \end{tabular}
\end{table*}

\subsection{Impact of Sequence Length}
To verify whether BITRec alleviates the ``semantic decoupling'' and ``length explosion'' issues in generative modeling, we evaluate all methods across sequence lengths $L \in \{25, 50, 75, 100\}$ on Taobao and Tmall (Figure~\ref{fig:ablation_study_sequence_length}).

\vspace{2pt}
\noindent\textbf{Resilience to Long-Sequence Degradation.}
As sequence length increases, generative methods (MBGen) and hierarchical models (HSTU) tend to plateau or degrade, particularly on Tmall---interleaving behaviors and items without explicit structure modeling makes long-range dependency capture increasingly difficult. Conversely, BITRec's advantage \textit{widens} as sequence lengthens: the MRR gap over SASRec on Taobao grows from 0.3 (L=25) to 0.4 (L=100). This confirms that joint embedding addresses semantic decoupling, while TRE's structural constraints prevent attention from getting lost in long, heterogeneous histories.

\subsection{Effectiveness of HBA Components}

Table~\ref{tab:ablation_hba_compact} presents ablation results for Hierarchical Behavior Aggregation (HBA). We design three variants: (1) \textbf{w/o HBA} removes the entire module; (2) \textbf{w/o Intensity Split} retains dual-channel architecture but treats all behaviors uniformly; (3) \textbf{Purchase-Only High} treats only purchase as high-intensity.

\vspace{2pt}
\noindent\textbf{Behavior Partition.} Following conversion funnel theory \cite{pi2019}, we partition behaviors by purchase intent: exploration behaviors (e.g., click, view) as low-intensity, commitment behaviors (e.g., favorite, cart, purchase) as high-intensity. For Insurance Dataset: $\mathcal{B}_{\text{low}}$=\{expose, click, search\}, $\mathcal{B}_{\text{high}}$=\{favorite, consult, purchase\}.

\vspace{2pt}
\noindent\textbf{Results Analysis.} Removing HBA causes consistent degradation across all datasets, with the largest drop on Insurance (up to 6.0\% NDCG@10)---aligning with domain characteristics where insurance decisions involve multi-stage deliberation. \textbf{w/o Intensity Split} underperforms the full model, validating that exploration and commitment behaviors encode distinct user states requiring separate aggregation. \textbf{Purchase-Only High} improves over uniform treatment but falls short of full partitioning, confirming that intermediate commitment signals (cart, favorite) also carry predictive value beyond purchase alone.

\subsection{Effectiveness of TRE Components}

Table~\ref{tab:ablation_tre} presents fine-grained ablations for Transition Relation Encoding (TRE). We evaluate four components: (1) \textbf{Global Behavioral Transition} captures learnable pairwise transition affinities between behavior types; (2) \textbf{Temporal Dynamics} encode multi-scale time decay; (3) \textbf{Item-Level Consistency} indicates product/category consistency; (4) \textbf{Implicit Context Matching} learns implicit context-dependent relevance.

\vspace{2pt}
\noindent\textbf{Component Contributions.} All components contribute positively with varying importance. Temporal Dynamics and Global Behavioral Transition yield the largest gains (1-2\% drops when removed). Item-Level Consistency shows dataset-dependent utility: minimal impact on Taobao where rapid browsing dominates, but meaningful on Insurance Dataset where same-product revisits signal deliberation.

\subsection{Training Efficiency}
Except for COPF, all methods maintain similar training costs on eight A100 GPUs: $\sim$2h (RetailRocket), $\sim$8h (Taobao), $\sim$14h (Tmall), and $\sim$1h (Insurance). COPF fails to converge within 10 days on Taobao and Tmall, while BITRec's HBA and TRE modules introduce only modest overhead.